\documentclass[a4paper]{jpconf}
\usepackage{graphicx}
\usepackage{amsmath}
\usepackage{amssymb}
\usepackage{hyperref}
\usepackage{subfigure}
\hypersetup{
    colorlinks = true,
    linkcolor=blue,
    anchorcolor=blue,
    urlcolor   = blue,
    citecolor  = blue,
}
\usepackage{multirow}
\usepackage{makecell}
\graphicspath{{./figs/}}

\begin{document}
\title{Prediction of turbulent energy based on low-rank resolvent modes and machine learning}

\author{Yitong Fan$^{1,2}$, Bo Chen$^{1}$, and Weipeng Li$^{1}$}
\address{$^1$ School of Aeronautics and Astronautics, Shanghai Jiao Tong University, Shanghai 200240, China}
\address{$^2$ Department of Mechanical Engineering, University of Melbourne, Parkville, VIC 3010, Australia}

\ead{liweipeng@sjtu.edu.cn}

\begin{abstract}
A  modelling framework based on the resolvent analysis and machine learning is proposed to predict the turbulent energy in incompressible channel flows. In the framework, the optimal resolvent response modes are selected as the basis functions modelling the low-rank behaviour of  high-dimensional nonlinear turbulent flow-fields, and the corresponding weight functions are determined by data-driven neural networks.
Turbulent-energy distribution in space and scales, at the friction Reynolds number 1000, is predicted and compared to the data of direct numerical simulation.
Close agreement is observed, suggesting the feasibility and reliability of the proposed framework for turbulence prediction.
\end{abstract}

\section{Introduction}
Understanding the wall-bounded turbulence is of great significance for the aerodynamic design and flow control in practical engineering applications.
To acquire precise spatial-temporal representations of turbulent motions, high-fidelity numerical simulations, such as direct numerical simulation (DNS) and wall-resolved large eddy simulation (WRLES), and three-dimensional particle image velocimetry (3D-PIV) have been widely performed in recent decades. However, these methods are  cost prohibitive. On the other hand, techniques such as Reynolds-averaged Navier-Stokes equations (RANS), hybrid RANS/LES, and low-cost experiments provide access to accurate mean flow fields, but cannot predict perturbations well.
Hence, relevant methods to efficiently predict the statistics and features of turbulent structures are highly desired. 

To reduce the complexity of the high-dimensional problem of turbulent flows, reduced-order models are often adopted as basis functions to estimate the full flow field, e.g. the widely used proper orthogonal decomposition (POD) \cite{Lumley1967,Berkooz1993}, spectral proper orthogonal decomposition (SPOD) \cite{Taira2017,Towne2018} and the dynamic mode decomposition (DMD) \cite{Schmid2010}. These methods are technically data-driven and usually require extensive post-processing \cite{Arun2023}. 
To enable the flow prediction in a physics-enhanced way, an equation-based method proposed from the perspective of linear dynamical system theory, named resolvent analysis \cite{McKeon2010}, is adopted in the present research. 
In this framework, an input-output system is obtained by linearizing Navier-Stokes equations around the steady base flow, with the input (forcing) and output (response) related by a linear resolvent operator. 
During the recent years, the resolvent analysis has been widely acknowledged in that it successfully provides a promising tool to identify the prominent linear mechanisms and predict the turbulent fluctuations and structures. 

Singular value decomposition is conducted upon the resolvent operator, to yield responses to the endogenous stimulus ranked by the amplification rate. 
The low-rank response modes are able to identify the the features in energy distribution and coherent structures, thus providing a promising tool to gain insight into the origins of the turbulent fluctuations and structures \cite{McKeon2017} and subsequently to enable the turbulence modelling and flow predictions. This is proved to be especially efficient in the incompressible wall-bounded turbulence.
Sharma et al. \cite{Sharma2013} identified the coherent structures in the turbulent pipe flow, including the velocity streaks and  packets of hairpin vortices, based on optimal response modes at representative wavenumber combinations.
They confirmed that a simple linear combination of optimal modes is able to constitute the ``skeleton'' of turbulence.
Abreu et al. \cite{Abreu2020} used resolvent analysis and SPOD to identify the dominant near-wall coherent structures in channel flows. 
Good agreement was observed between the low-rank representations of SPOD and resolvent analysis, especially at the frequency and wavenumbers where the lift-up mechanism is present. 
Moarref et al. \cite{Moarref2013} exploited the optimal resolvent modes, subject to the non-broadband forcing in wave speed, to predict the streamwise velocity fluctuations in high-Reynolds-number channel flows. With a positive weight function amplifying or attenuating the streamwise energy density of the optimal mode, they solved a convex optimization problem so as to match the model-based and the direct-numerical-simulation(DNS)-resolved streamwise velocity fluctuations. Similar methods were applied to estimate the distribution of spanwise and wall-normal velocity fluctuations and the Reynolds shear stress across the wall layer \cite{Moarref2013a, Moarref2014}. Further investigations on the influence of the selected numbers of low-rank modes on the prediction capability were conducted by Moarref et al. \cite{Moarref2014}. They found that optimally forced (rank-1) response mode is able to well capture the streamwise energy. However, to estimate the spanwise and wall-normal energy spectrum and the streamwise-wall-normal co-spectrum, some more higher-order modes might be in necessity. 
Symon et al. \cite{Symon2021} predicted the evolution of turbulence kinetic energy, in plane Poiseuille flows, based on the optimal response modes of velocity fluctuations from resolvent analysis, with the addition of eddy viscosity.
The successful usage of the resolvent operator in the fully developed turbulence helps to identify the equation-based models for prediction.

In addition, more other strategies for flow estimation/reconstruction on the basis of resolvent framework have emerged in the recent years, e.g. constructing a suitable transfer function from the measurements to the driving force \cite{Towne2020,Amaral2021,Illingworth2018,Martini2020, Arun2023} and appropriately shaping the nonlinear forcing terms in the input-output framework \cite{Morra2019,Morra2021,Holford2023}. However, these methods are usually used case by case.
To enable the resolvent-based prediction in a computationally feasible and generalizable way, we intend to employ data-driven methods in combination with the resolvent analysis. Recent advances in machine learning have shown great potential for a fast alternative method in fluid mechanics \cite{Brunton2020}, due to its inherent strengths in managing high-dimensional data compression, feature extraction, and generative inversion \cite{Kou2021}. Notwithstanding, the interpretability of purely data-driven methods is highly restricted and the training of these models is reliant on a sufficient amount of data.
Studies have been seeking to address this issue by integrating existing knowledge of physics with machine learning methods, yielding satisfying outcomes \cite{physics_enhanced, physics_constrain,Raissi2019, Cai2021}.

In the present study, we intend to exploit the resolvent analysis to obtain a low-rank approximation of the nonlinear system as basis functions, which is characteristic of the turbulence cycle from a physics-enhanced perspective. 
Network models are then reconstructed to predict high-dimensional coefficients of the low-rank forcing modes in the wavenumber-frequency domain, with a set of mean velocity profiles being the input. Linear combination of resolvent response modes and the coefficient matrix, i.e. a weighted sum of the resolvent modes, consequently leads the reconstruction of three-dimensional flow fields. Using the limited database of turbulent channel flows, this paper targets at predicting the wall-normal distribution of turbulent energy and the premultiplied spectra of turbulent fluctuations.
The remainder of this paper is outlined as follows. Section \ref{method} introduces the resolvent formulation and the machine-learning framework. Databases that are used for training and testing are also described. In Sec. \ref{result}, validation and verification of the resolvent-based reduced-order model is investigated and the the predicted results of the turbulent energy are shown, in comparison with the true results. Finally, concluding remarks are given in Sec. \ref{conclusion}.

\section{Methodology} \label{method}
The non-dimensionalized Navier-Stokes equations for the incompressible turbulent channel flows are 
\begin{align}
\frac{\partial \boldsymbol{u}}{\partial t} + \boldsymbol{u}\cdot \nabla \boldsymbol{u}  &= - \nabla p + \frac{1}{Re} \nabla^2 \boldsymbol{u},\label{ns1}\\
\nabla \cdot \boldsymbol{u} &=0,
\label{ns2}
\end{align} 
where $t$ is time, $p$ is pressure, the vector $\boldsymbol{u}=[u,v,w]^\top$ contains the velocity components in the streamwise ($x$), wall-normal ($y$) and spanwise ($z$) directions, and $\nabla = [\partial/\partial x, \partial/\partial y, \partial/\partial z]^\top$.
The bulk Reynolds number $Re$ is defined by $u_b^* h^*/\nu^*$, with $u_b$ being the bulk velocity, $h$ the channel half height, and $\nu$ the kinematic viscosity. The superscript $*$ denotes the dimensional values, otherwise the quantities are non-dimensionalized.

\subsection{Resolvent formulation}

Consider a  state variable $\boldsymbol{q} = [u,v,w,p]^\top$ in the statistically steady turbulent channel flows. It can be Fourier transformed in the temporal and homogenous spatial (streamwise and spanwise) directions,
\begin{equation}
\boldsymbol{q}\left(x, y, z, t\right)=\iiint_{-\infty}^{\infty} \hat{\boldsymbol{q}}\left(y ; \kappa_{x}, \kappa_{z}, \omega\right) \mathrm{e}^{\mathrm{i}\left(\kappa_{x} x+\kappa_{z} z-\omega t\right)} \mathrm{d} \kappa_{x} \mathrm{d} \kappa_{z} \mathrm{d} \omega.
\label{ft}
\end{equation}
The $\hat{(\cdot)}$ denotes the Fourier-transformed variables, $\kappa_x$ and $\kappa_z$ are the streamwise and spanwise wavenumbers, $\omega$ is the temporal frequency, and $\mathrm{i}=\sqrt{-1}$. The relevant wavelengths in the streamwise and spanwise directions are defined by $\lambda_x = 2\pi/\kappa_x$ and $\lambda_z = 2\pi/\kappa_z$, respectively. The mean state of $\boldsymbol{q}$ varies only in the wall-normal direction, i.e. $\bar{\boldsymbol{q}}(y) = [\bar{u}(y),0,0,\bar{p}(y)]^\top$, identical to the $\hat{\boldsymbol{q}}$ at $(\kappa_x, \kappa_z,  \omega)= (0,0,0)$.

Substituting \eqref{ft} into the Navier-Stokes equations \eqref{ns1} and \eqref{ns2}, they are recast in the Fourier form for each  $(\kappa_x, \kappa_z,  \omega)\ne(0,0,0)$ as:
\begin{equation}
-\mathrm{i} \omega \hat{\boldsymbol{q}}\left(y ; \kappa_{x}, \kappa_{z}, \omega\right)  = \boldsymbol{A}\left(\kappa_{x}, \kappa_{z}, \omega\right) \hat{\boldsymbol{q}}\left(y ; \kappa_{x}, \kappa_{z}, \omega\right)  +\boldsymbol{B}\left(\kappa_{x}, \kappa_{z}, \omega\right) \hat{\boldsymbol{f}}\left(y ; \kappa_{x}, \kappa_{z}, \omega\right).
\end{equation}
The $\hat{\boldsymbol{f}}$ is the input forcing containing all the nonlinear contributions. It is comprised of streamwise, wall-normal and spanwise components, i.e. $\hat{\boldsymbol{f}}=[f_x,f_y,f_z]^\top$. $\boldsymbol{A}$ is a linear operator and $\boldsymbol{B}$ is the input matrix restricting the forcing terms to exist only in the momentum equations. 

It consequently leads to a linearized form of the governing equations:
\begin{align}
\hat{\boldsymbol{q}}\left(y ; \kappa_{x}, \kappa_{z}, \omega\right) &= \left(-\mathrm{i} \omega \boldsymbol{I}-\boldsymbol{A}\left(\kappa_{x}, \kappa_{z}, \omega\right)\right)^{-1} \boldsymbol{B}\left(\kappa_{x}, \kappa_{z}, \omega\right) \hat{\boldsymbol{f}}\left(y ; \kappa_{x}, \kappa_{z}, \omega\right) \nonumber \\
&=\boldsymbol{H}\left(\kappa_{x}, \kappa_{z}, \omega\right) \hat{\boldsymbol{f}}\left(y ; \kappa_{x}, \kappa_{z}, \omega\right).
\label{resolvent}
\end{align}
$\boldsymbol{I}$ is the identity matrix, and $\boldsymbol{H} = (-\mathrm{i} \omega \boldsymbol{I}-\boldsymbol{A})^{-1} \boldsymbol{B}= \boldsymbol{L}^{-1} \boldsymbol{B}$ is the linear operator relating the input forcing to the output state. To simplify the notations, the variable dependency on $\left(y ; \kappa_{x}, \kappa_{z}, \omega\right)$ will be dropped hereafter.
The expressions of $\boldsymbol{L}$ and $\boldsymbol{B}$ are given by
\begin{align}
\boldsymbol{L}=\left[\begin{array}{cccc}
\mathrm{i} k_x \bar{\boldsymbol{u}}-\mathrm{i} \omega \boldsymbol{I}-\frac{1}{R e} \nabla^2 & \frac{d \bar{\boldsymbol{u}}}{d \boldsymbol{y}} & \mathbf{0} & \mathrm{i} k_x \boldsymbol{I} \\
\mathbf{0} & \mathrm{i} k_x \bar{\boldsymbol{u}}-\mathrm{i} \omega \boldsymbol{I}-\frac{1}{R e} \nabla^2 & \mathbf{0} & \frac{d}{d \boldsymbol{y}} \\
\mathbf{0} & \mathbf{0} & \mathrm{i} k_x\bar{\boldsymbol{u}}-\mathrm{i} \omega \boldsymbol{I}-\frac{1}{Re} \nabla^2 & \mathrm{i} k_z \boldsymbol{I} \\
\mathrm{i} k_x \boldsymbol{I} & \frac{d}{d \boldsymbol{y}} & \mathrm{i} k_z \boldsymbol{I} & \mathbf{0}
\end{array}\right],
\label{eq:L}
\end{align}

\begin{align}
\boldsymbol{B}=\left[\begin{array}{ccc}
\boldsymbol{I} & \boldsymbol{0} & \boldsymbol{0} \\
\boldsymbol{0} & \boldsymbol{I} & \boldsymbol{0} \\
\boldsymbol{0} & \boldsymbol{0} & \boldsymbol{I} \\
\boldsymbol{0} & \boldsymbol{0} & \boldsymbol{0}
\end{array}\right], 
\end{align}
where $\nabla^2=d^2 / d \boldsymbol{y}^2-(k_x^2+k_z^2) \boldsymbol{I}$. 

In the resolvent formulation \eqref{resolvent}, even though the nonlinear forcing is unknown, the dominant characteristics of turbulent fluctuations can be captured by the low-rank modes of $\boldsymbol{H} $, as is confirmed by various previous studies \cite{McKeon2010,Abreu2020,Moarref2013}. Before we proceed to apply singular value decomposition (SVD) to the resolvent operator, we introduce the definition of the inner product utilized herein, which is able to characterize the turbulent motions from the perspective of turbulent kinetic energy of the whole dynamical system. Let
\begin{equation}
\left<\hat{\boldsymbol{q}},\hat{\boldsymbol{q}}\right>=\hat{\boldsymbol{q}}^\dag \boldsymbol{W_q}\hat{\boldsymbol{q}},
~\left<\hat{\boldsymbol{f}},\hat{\boldsymbol{f}}\right>=\hat{\boldsymbol{f}}^\dag \boldsymbol{W_f}\hat{\boldsymbol{f}},~
\end{equation}
where the superscript $\dag$ denotes the complex conjugate transpose. The weight matrices are given by 
\begin{align}
\boldsymbol{W_q}=\left[\begin{array}{cccc}
\boldsymbol{K} & \boldsymbol{0} & \boldsymbol{0} & \boldsymbol{0}\\
\boldsymbol{0} & \boldsymbol{K} & \boldsymbol{0} & \boldsymbol{0}\\
\boldsymbol{0} & \boldsymbol{0} & \boldsymbol{K} & \boldsymbol{0}\\
\boldsymbol{0} & \boldsymbol{0} & \boldsymbol{0} & \boldsymbol{0}
\end{array}\right]\quad
{\rm and}\quad
\boldsymbol{W_f}=\left[\begin{array}{ccc}
\boldsymbol{K} & \boldsymbol{0} & \boldsymbol{0} \\
\boldsymbol{0} & \boldsymbol{K} & \boldsymbol{0} \\
\boldsymbol{0} & \boldsymbol{0} & \boldsymbol{K}
\end{array}\right].
\end{align}
The $\boldsymbol{K}$ is the Clenshaw-Curtis quadrature weights \cite{Trefethen2000} in the wall-normal direction. Taking the $\boldsymbol{W_q}$ and $\boldsymbol{W_f}$ into consideration, we perform SVD on the corrected resolvent operator to extract the low-rank approximations, that is,
\begin{align}
\boldsymbol{W_q}^{1/2}\boldsymbol{H}\boldsymbol{W_f}^{-1/2} = \sum\limits_{p=1}^{N}\tilde{\boldsymbol{\psi}}_p \sigma_p \tilde{\boldsymbol{\phi}}_p^\dag,\\
\boldsymbol{\phi}_p =\boldsymbol{W_f}^{-1/2} \tilde{\boldsymbol{\phi}}_p,\\
\boldsymbol{\psi}_p =\boldsymbol{W_q}^{-1/2} \tilde{\boldsymbol{\psi}}_p,
\end{align}
where $N$ is the number of resolvent modes, $\boldsymbol{\psi}_p$ and $\boldsymbol{\phi}_p$ are $p$th-rank orthogonal basis functions of the response and forcing modes. The singular value $\sigma_p$ denotes the energy amplification rate, ranked by $\sigma_{p}>\sigma_{p+1}$. Based on this modal decomposition, the nonlinear terms and the state variables can be reconstructed from a weighted sum of all the forcing and response modes:
\begin{align}
\hat{\boldsymbol{f}}&=\sum\limits_{p=1}^{N}{a}_p \boldsymbol{\phi}_p,\\
\hat{\boldsymbol{q}}&=\sum\limits_{p=1}^{N}\boldsymbol{\psi}_p \sigma_p a_p.
\label{qall}
\end{align}
The $a_p$ is the projection of the forcing mode onto the nonlinear term.

\subsection{Prediction of turbulent energy based on low-rank resolvent modes and machine learning} \label{predictionmethod}
In wall turbulence, the low-rank nature of the resolvent operator is widely observed, indicating that the understanding of the origins of the turbulent fluctuations and structures can be realized using the rank-1 (optimal) model. Therefore, in the present study, we majorly consider this simple response mode as a basis function of the turbulent fields, which can be recovered with a linear superposition of optimal modes at all wavenumber combinations. 
The premultiplied spectra of turbulent fluctuations carried by the optimal mode can be obtained by 
\begin{equation}
E_{qq}\left(y ; \kappa_x, \kappa_z, c \right)=\kappa_x^2 \kappa_z\left(|\boldsymbol{\psi}_1|\left(y ; \kappa_x, \kappa_z, c\right) \sigma_1\left( \kappa_x, \kappa_z, c \right) a_1\left( \kappa_x, \kappa_z, c \right)\right)^2.
\label{Euu}
\end{equation}
The $\omega-$dependence is replaced by the wave speed $c$, expressed as $c=\omega/\kappa_x$, according to the grid settings of the resolvent operator which will be  accounted for in the following.
Integrating \eqref{Euu} over the frequency and spatial wavenumber domain yields the one-dimensional premultiplied spectra and the turbulent energy density in the wall-normal direction, viz.
\begin{align}
E_{qq}(y, \kappa_x)  &=\iint E_{qq}\left(y ; \kappa_x, \kappa_z, c\right) \mathrm{d} \log \left(\kappa_z\right) \mathrm{d} c, \\
E_{qq}(y, \kappa_z)  &=\iint E_{qq}\left(y ; \kappa_x, \kappa_z, c\right) \mathrm{d} \log \left(\kappa_x\right) \mathrm{d} c, \\
E_{qq}(y)  &= \iiint E_{qq}\left(y ; \kappa_x, \kappa_z, c\right) \mathrm{d} \log \left(\kappa_x\right) \mathrm{d} \log \left(\kappa_z\right) \mathrm{d} c.
\end{align}

In this framework, to predict the $E_{qq}(y, \kappa_x)$, $E_{qq}(y, \kappa_z)$ or $E_{qq}(y)$,  the projection coefficients of the resolvent modes onto the real forcing contribution, i.e. the $a_1$ in \eqref{Euu}, are of great significance and to be determined. To quantify $a_1$ in a full field of spatial wavenumber and frequency pairs in a computationally feasible and generalizable way, we intend to employ data-driven methods herein. 

\begin{figure}
\centering
\includegraphics[width=1.\textwidth]{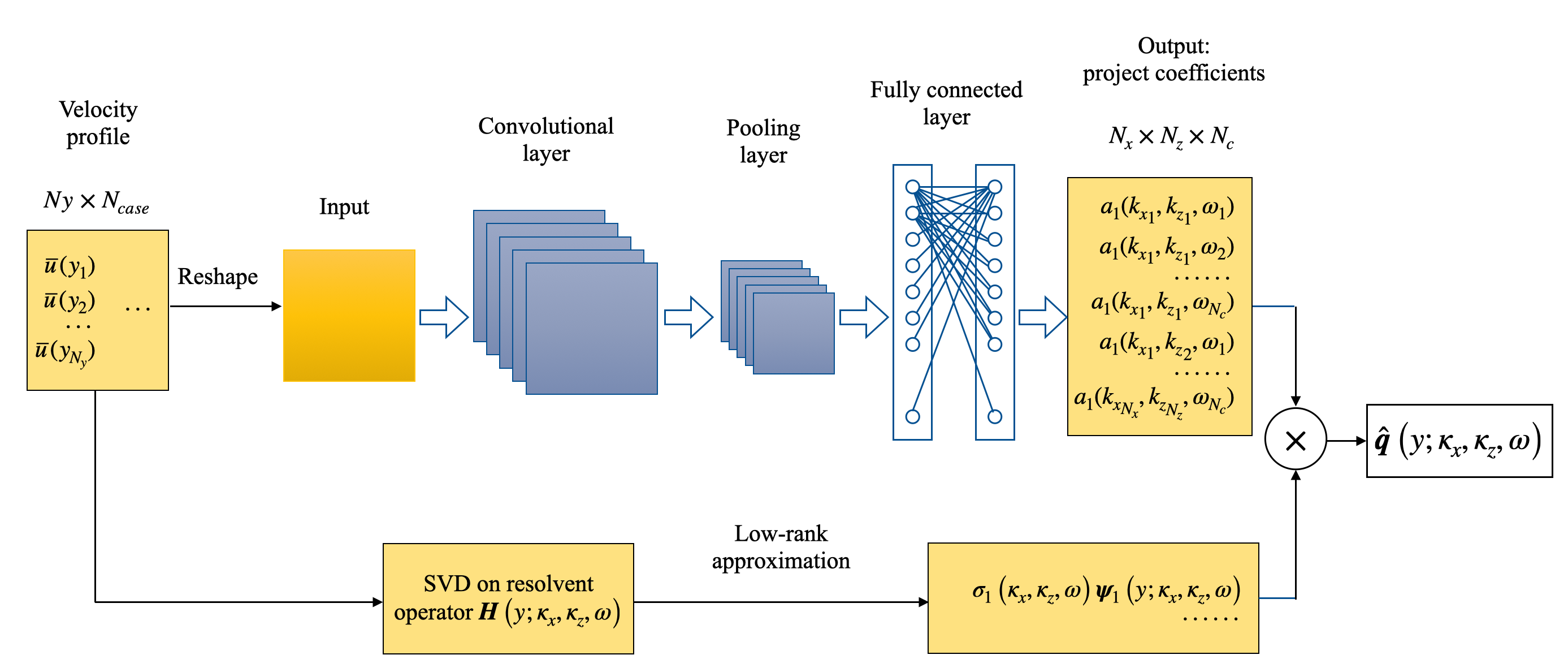}
\caption{The schematic of the network for prediction of the projection coefficients.}\label{network}
\end{figure}

The training network models have been developed, working as a projection function of the input, shown in figure \ref{network}. The cases of incompressible turbulent channel flows at various Reynolds numbers are considered. Since only the mean streamwise velocity is required in the resolvent operator (see equation \eqref{eq:L}), the mean velocity profiles across the wall layer is fed to the network.
In figure \ref{network}, we use the architecture of convolutional neural network (CNN), which is especially designed for large-scale structured data. The CNN is mainly composed of convolutional layers, pooling layers, and fully connected layers. The convolutional layer  captures the global and local information of the input data, and the pooling layer aims to reduce the dimension of the data to avoid overfitting. 
We reshape the mean profile in the turbulent boundary layers to satisfy the requirement of input, and make the full-field projection coefficients to be the output. In this way, turbulent energy $E_{qq}$ can be consequently obtained following the low-rank approximation of the resolvent analysis. 
Beyond that, with the inverse transform, this framework is a quite promising tool to quantify instantaneous three-dimensional flow field as shown in figure \ref{network}, as long as sufficient databases are exploited for training and testing.
This is out of the scope of this paper's target.

\subsection{Databases and computational methods}
Incompressible turbulent channel flows, at the friction Reynolds numbers $Re_\tau=180,390,550,1000,2000,4200$ and $5200$, are considered in the present study. The mean base flow for input and the energy density for prediction is obtained from the direct numerical simulations in Refs. \cite{Moser1999,LozanoDuran2014,Lee2015}.

To obtain the resolvent modes at all required frequency and spatial wavenumber pairs, a sufficient domain with $N_x \times N_z \times N_c=87\times87\times24$ points is constructed. $N_x$ and $N_z$ are the numbers of grid points for the streamwise and spanwise wavelength, which are logarithmically spaced with the grid resolutions being $\Delta \log(\lambda_x)=0.1$ and $\Delta \log(\lambda_z)=0.1$, for all the cases. In the frequency domain, $N_c$ denotes the number of grid points for the streamwise wave traveling at the speed of $c$. Since it has been found that the wave speeds are energetically important within $2 \le c/u_\tau \le \bar{u}_{c}/u_\tau$ \cite{Moarref2013}, where the $\bar{u}_{c}$ is the mean velocity at the channel centerline and $u_\tau$ is the friction velocity, the wave speeds can be linearly chosen in this range to efficiently reduce the computational cost, and the excluded wave parameters will not influence the following results. 
In the wall-normal direction, $N_y$ grid points are spaced via the Chebyshev collocation method, defined by $y=\cos (\pi j/(N_y-1))$, where $j=0,1,...,N_y-1$ and $-1 \le y \le 1$. In the present study, we make $N_y=200$ for all the cases and the second-layer inner-scaled height $y_{min}^+$ restricted within 0.65. The non-slip impermeable boundary condition is applied at the wall.

\section{Results and discussion} \label{result}
In this section, we will first extract the optimal response modes in channel flows, evaluating their capability in identifying the coherent structures and recovering the energy distribution. The projection coefficients of the low-rank forcing modes are then obtained through machine learning, adjusting the magnitude of the response, to predict the spectra and wall-normal distributions of turbulent energy. 

\subsection{Low-rank modelling of the coherent structures}
\begin{figure}[hpbt!]
\centering
\subfigure{\includegraphics[width=.4\textwidth]{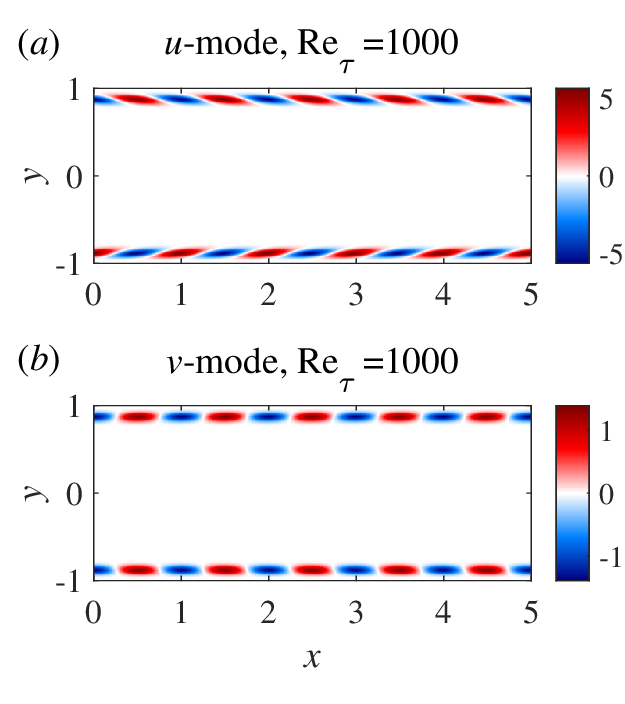}}\subfigure{\includegraphics[width=.6\textwidth]{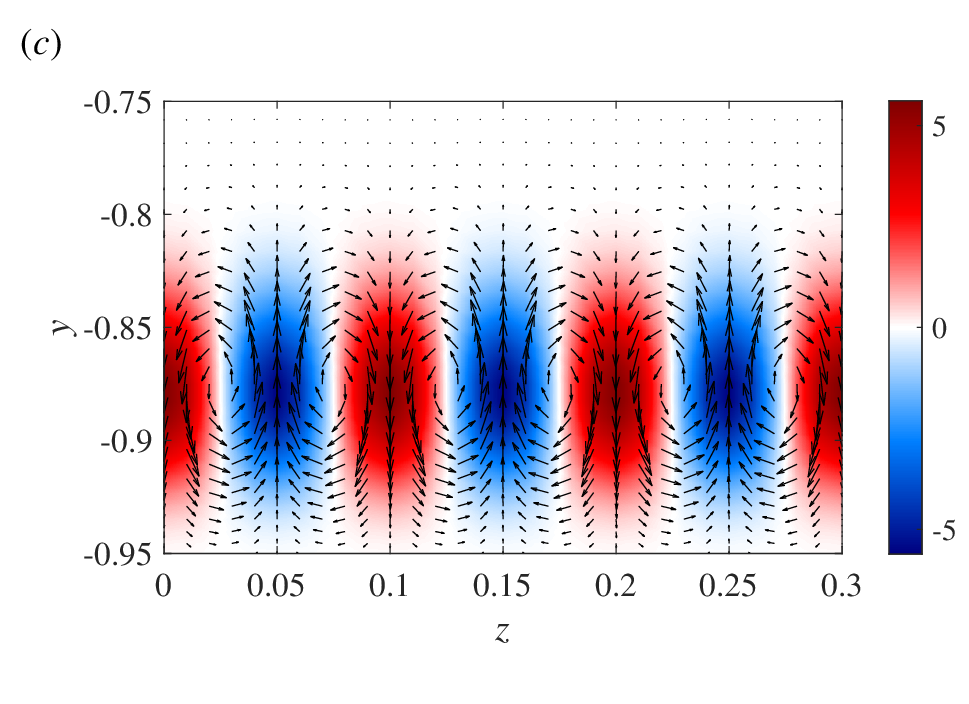}}
\caption{The response mode shapes of the ($a$,$b$) streamwise and wall-normal velocity in the streamwise-wall-normal plane, and ($c$) streamwise velocity in the cross-stream plane at $(\lambda_x^+,\lambda_z^+,c^+)=(1000,100,16.8)$, in the channel flow at $Re_\tau=1000$. Red and blue denote positive and negative disturbances, respectively. The arrows denote the wall-normal and spanwise velocity components of the vortices.}\label{shape}
\end{figure}

Figure \ref{shape} shows the shapes of the velocity and temperature mode  at a representative wavenumber combination $(\lambda_x^+,\lambda_z^+,c^+)=(1000,100,16.8)$. The superscript $+$ denotes the normalization with inner units. 
Only the results in the channel flow at $Re_\tau=1000$ are shown here for brevity, and those in the other cases do not affect the following conclusion.  In panels ($a$) and ($b$), positive and negative velocity fluctuations appear alternatively in the streamwise direction, with the $u$-mode attached to the wall with a narrow angle while the $v$-mode without an inclination angle to the wall. This is associated with the mild phase variation of $v$-response in the wall-normal direction. The signs of the $u$- and $v$-mode are negatively correlated above the bottom surface, meaning that the low- or high-speed streaks (denoted by the negative or positive $u$-fluctuations) are carried away from or towards  the wall. This phenomenon is more straightforwardly depicted in the cross-stream plane, shown in panel ($c$). The high- and low-speed streaks accompanied by streamwise vortices expose the sweep and ejection events, consistent with the lift-up mechanism, capturing the empirically observed characteristics of the near-wall coherent structures in the wall-bounded turbulence.

\begin{figure}[hpbt!]
\centering
\subfigure{\includegraphics[width=.5\textwidth]{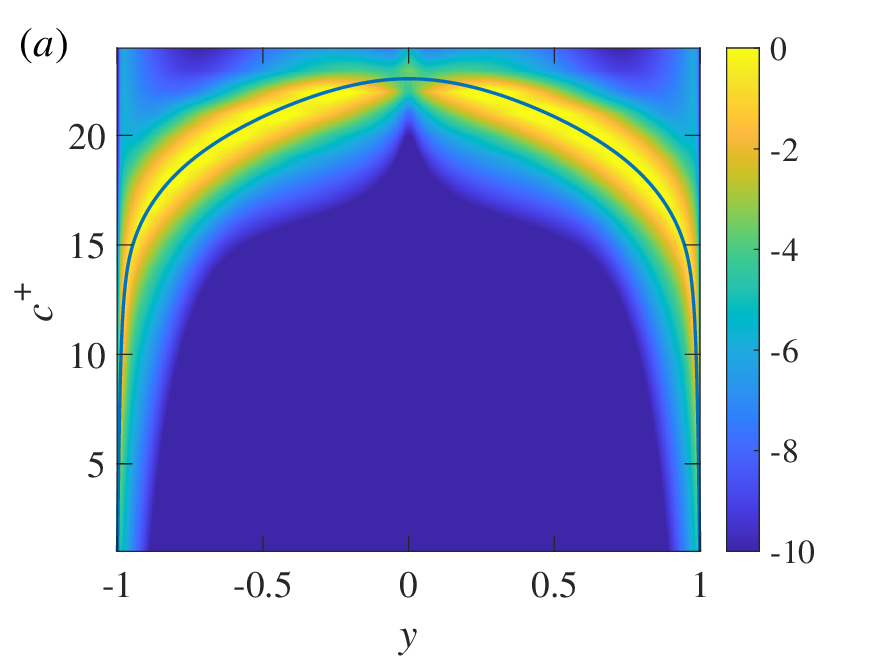}\label{energy:a}}\subfigure{\includegraphics[width=.5\textwidth]{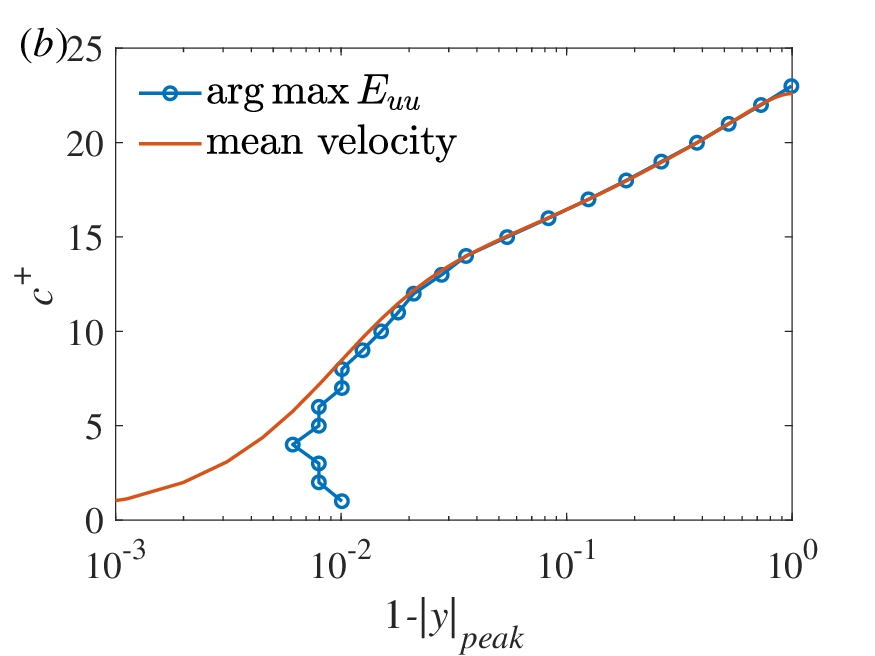}\label{energy:b}}
\caption{($a$) Premultiplied one-dimensional energy spectra of the streamwise viscosity, and ($b$) the wall-normal distance of the peak streamwise velocity amplitude as a function of the wave speed, in the channel flow at $Re_\tau=1000$. The solid lines in panel ($a$) and ($b$) represent the mean streamwise velocity profile issued from DNS.}\label{energy}
\end{figure}

Figure \ref{energy} shows the premultiplied one-dimensional energy spectra in the optimal streamwise response, given by 
\begin{equation}
E_{uu}=\frac{\iint_{-\infty}^\infty k_x^2k_z(\sigma_1 |\psi_{1,u}|)^2 \mathrm{d} \log \left(\kappa_x\right) \mathrm{d} \log \left(\kappa_z\right)}{{\rm max} \iint_{-\infty}^\infty k_x^2k_z\sigma_1^2 \mathrm{d} \log \left(\kappa_x\right) \mathrm{d} \log \left(\kappa_z\right)},
\end{equation}
where $\psi_{1,u}$ is the streamwise-velocity component of the optimal response mode, and the forcing is simply assumed to be a white-noise signal. In figure \ref{energy:a}, the premultiplied spectra are plotted in the logarithmic scale.
It is observed that the energy density is clustered around the mean velocity profile, suggesting that the response mode follows  the Taylor's frozen-turbulence hypothesis, in the sense of critical-layer mechanism \cite{McKeon2010}. 

The wall-normal distance of peak streamwise velocity amplitude is quantified as a function of the wave speed $c^+$ in figure \ref{energy:b}.
The most energetic wave-propagating speed profile is well collapsed onto that of the mean streamwise velocity, which is consistent with the critical-layer observation in figure \ref{energy:a}. Hence the corresponding mode, which travels at the speed $c^+$ and is located at the wall-normal distance $y_{peak}$, is termed the ``critical response mode''. 
Exceptions are observed in the very-near-wall region, e.g. $y^+<10$ ($y^+=(1-|y|)/\delta_\nu$ with $\delta_\nu$ being the viscous length scale), where the $y_{peak}-c^+$ distribution deviates from the mean velocity profile.
At this state, the streamwise turbulent energy is finite reaching down to the wall, suggesting that the mode has a footprint down to the wall. In this sense, it is termed  the ``wall-attached response mode'' \cite{McKeon2013}.
Overall, the phenomenon observed here in the optimal response mode is confirmed to be consistent with the classical turbulent structures and energy distribution in wall-bounded flows, hence it is rational to make it as a basis function for predicting the turbulent statistics.

\subsection{Prediction of the turbulent energy}

\begin{figure}
\centering
\subfigure{\includegraphics[width=1.\textwidth]{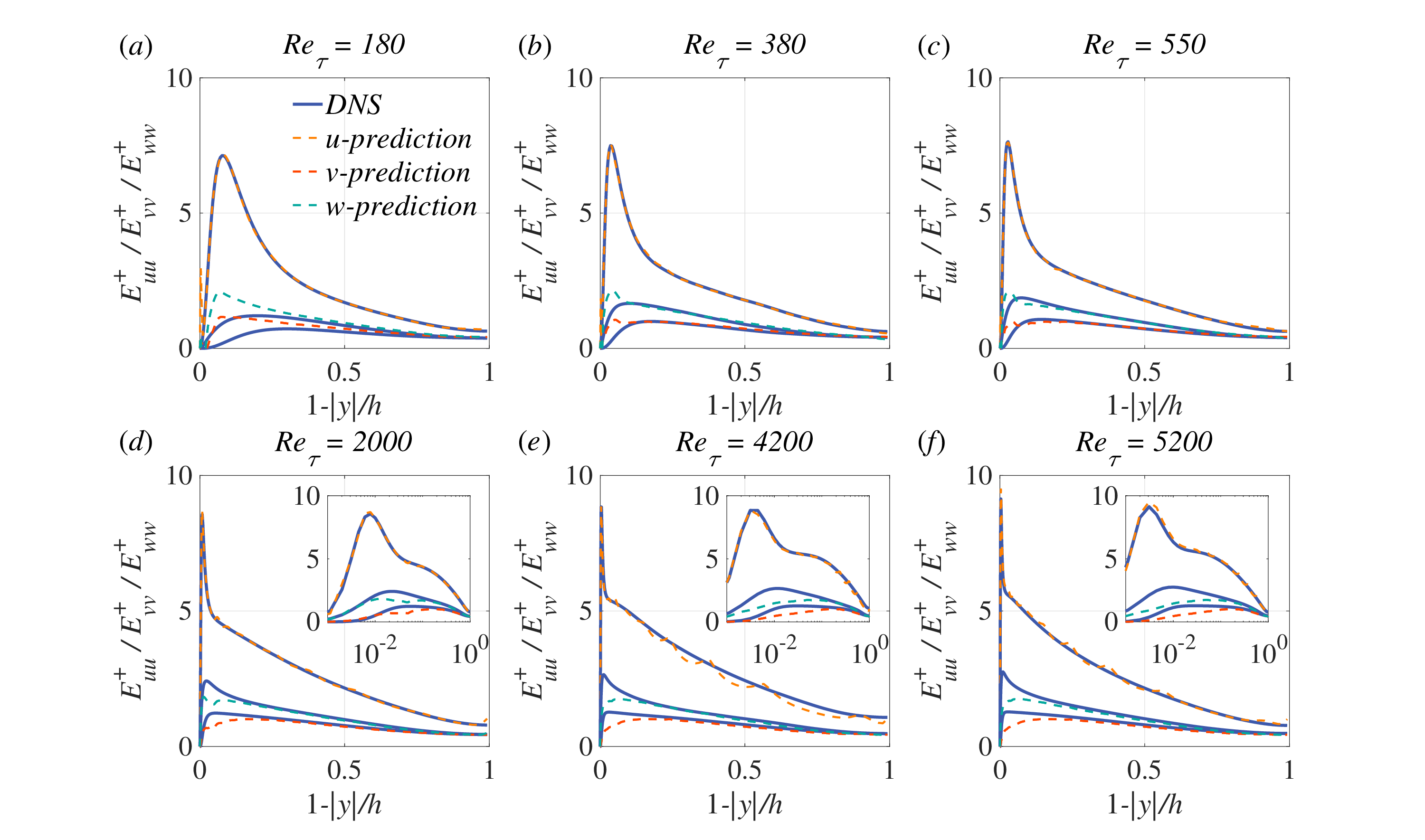}}
\caption{The prediction of $E_{uu}^+$, $E_{vv}^+$ and $E_{ww}^+$ as a function of wall-normal distance in the training process, when $(\gamma_{uu}, \gamma_{vv}, \gamma_{ww}, \gamma_{pp})=(1,0,0,0)$, in channel flows at various Reynolds numbers. The DNS results are also shown for comparison. Profiles are also plotted in logarithmic scale in insets of panels ($d$-$f$), to clarify the inner-layer behaviour at high Reynolds numbers. Legends in panels ($b$-$f$) refer to ($a$).}\label{training}
\end{figure}

\begin{figure}
\centering
\subfigure{\includegraphics[width=.9\textwidth]{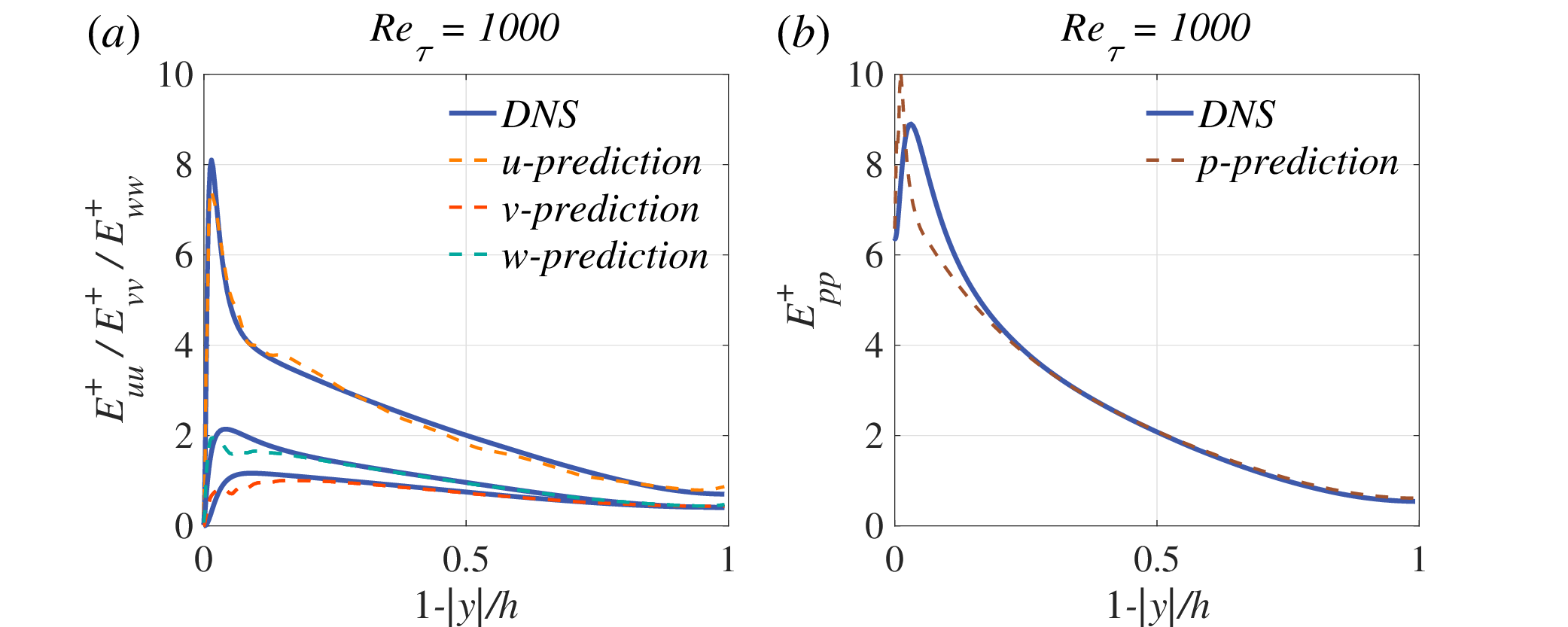}}
\caption{The prediction of ($a$) $E_{uu}^+$, $E_{vv}^+$ and $E_{ww}^+$ and  ($b$) $E_{pp}^+$ as a function of wall-normal distance in the testing process, when $(\gamma_{uu}, \gamma_{vv}, \gamma_{ww}, \gamma_{pp})=(1,0,0,0)$, in the channel flow at $Re_\tau=1000$.}\label{testing1}
\end{figure}

\begin{figure}
\centering
\subfigure{\includegraphics[width=.9\textwidth]{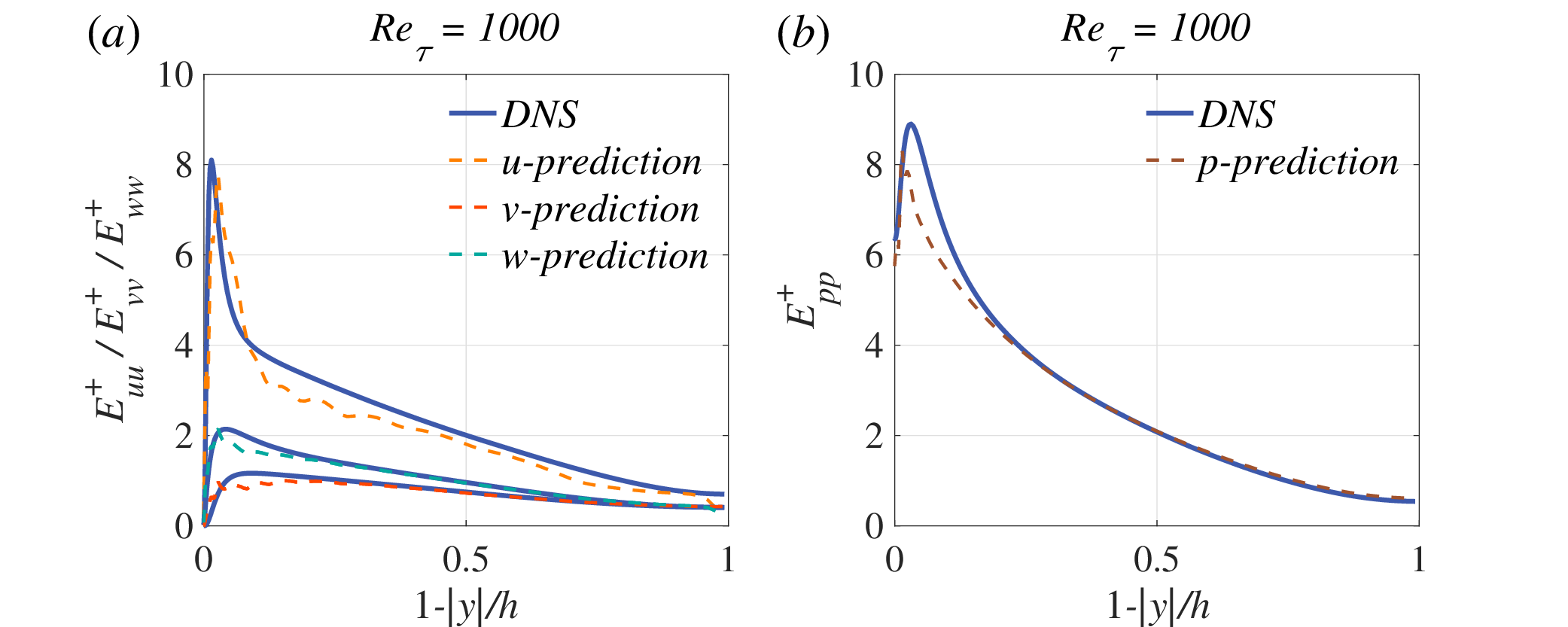}}
\caption{The prediction of ($a$) $E_{uu}^+$, $E_{vv}^+$ and $E_{ww}^+$ and  ($b$) $E_{pp}^+$ as a function of wall-normal distance in the testing process, when $(\gamma_{uu}, \gamma_{vv}, \gamma_{ww}, \gamma_{pp})=(1,1,1,1)$, in the channel flow at $Re_\tau=1000$.}\label{testing2}
\end{figure}

As introduced in Sec. \ref{predictionmethod}, this section aims to predict the turbulent energy in space and scales, by obtaining the $a_1$ as a function of $(k_x, k_z, c)$.
In the practical process of training, this would be rather difficult with the limited training databases. 
Hence the spatial broadband-forcing assumption is exploited, which means the dependency dimension of  $a_1$ is reduced, to enable the prediction and generalization of the full-field projection coefficients at a lower computational cost and memory requirement. For example, the spanwise premultiplied spectral density and the wall-normal distribution of the turbulent energy are rewritten, respectively, as
\begin{align}
E_{qq}(y, \kappa_z)  =\int a_1^2\left(\kappa_z, c \right) \left[\int \kappa_x^2 \kappa_z
|\boldsymbol{\psi}_1|^2\left(y ; \kappa_x, \kappa_z, c\right) \sigma_1^2\left( \kappa_x, \kappa_z, c \right) \mathrm{d} \log \left(\kappa_x\right) \right] \mathrm{d} c, \label{Eqq1}\\
E_{qq}(y)  =\int a_1^2\left(c \right) \left[\iint \kappa_x^2 \kappa_z
|\boldsymbol{\psi}_1|^2\left(y ; \kappa_x, \kappa_z, c\right) \sigma_1^2\left( \kappa_x, \kappa_z, c \right) \mathrm{d} \log \left(\kappa_x\right)  \mathrm{d} \log \left(\kappa_z\right) \right] \mathrm{d} c.
\end{align}
It has been confirmed that this reduced-order $a_1$ is still able to provide sufficient degrees of freedom for prediction \cite{Moarref2013}.

In the present study, a CNN architecture is employed, as shown in figure \ref{network}. To optimize the weight and bias parameters in neurons per hidden layer during the training process, the mean squared error (MSE) is defined by a loss function:
\begin{align}
MSE=&\gamma_{uu} \frac{||E_{uu,model}-E_{uu,true}||^2}{||E_{uu,true}||^2}+\gamma_{vv} \frac{||E_{vv,model}-E_{vv,true}||^2}{||E_{vv,true}||^2}\nonumber\\
+&\gamma_{ww} \frac{||E_{ww,model}-E_{ww,true}||^2}{||E_{ww,true}||^2}+\gamma_{pp} \frac{||E_{pp,model}-E_{pp,true}||^2}{||E_{pp,true}||^2},
\end{align}
where the $||\cdot ||$ represents the $L_2$ norm, and the subscript ``model'' and ``true'' denote the result from the resolvent analysis and DNS, respectively. The $\gamma_{uu}$, $\gamma_{vv}$, $\gamma_{ww}$ and $\gamma_{pp}$ controls different forms of the matching errors. 
Two kinds of loss functions are considered in the present study to train the network, that is (i) only the matching error of the streamwise energy is penalized, i.e. $(\gamma_{uu}, \gamma_{vv}, \gamma_{ww}, \gamma_{pp})=(1,0,0,0)$; and (ii) the matching error of all the four state variables is penalized, i.e. $(\gamma_{uu}, \gamma_{vv}, \gamma_{ww}, \gamma_{pp})=(1,1,1,1)$.

First, we present the training and testing performance in the prediction of the wall-normal distribution of energy, with six cases at $Re_\tau=180,380,550,2000,4200,5200$ used for  training, and one case at $Re_\tau=1000$ for testing.
Figure \ref{training} shows the output of streamwise ($E_{uu}^+$), wall-normal ($E_{vv}^+$) and spanwise energy ($E_{ww}^+$) in inner units, in comparison to the DNS results, in the training process with $(\gamma_{uu}, \gamma_{vv}, \gamma_{ww}, \gamma_{pp})=(1,0,0,0)$. It is observed that the predicted streamwise energy is almost collapsed onto the DNS profiles, especially at $Re_\tau \lesssim 2000$, confirming that the optimization process has been well converged. At higher Reynolds numbers, wiggles in $E_{uu}^+$ are observed in the outer region, which is probably associated with the limited wall-normal grid points. Note that we make $N_y=200$ for all the cases under scrutiny, which is both computationally feasible and generally sufficient to discretize the differential operators in the wall-normal direction.

Whereas for the prediction of wall-normal and spanwise component, obvious deviations are observed in contrast to the DNS results, which is partly due to that only the matching error of streamwise energy is considered in the training process. 
For channel flows, the streamwise velocity dominates the other velocity components, leading to significant implications on projecting the resolvent modes onto the real distributions issued from DNS in the streamwise direction. Since the inherent ratio of the streamwise response to the wall-normal and spanwise response in the rank-1 model does not match that in the DNS results, the same projection coefficients that optimally recover $E_{uu}^+$ cannot perfectly yield results of $E_{vv}^+$ and $E_{ww}^+$.
The corresponding energy prediction in the testing process, where $Re_\tau=1000$, is depicted in figure \ref{testing1}.   Similar phenomenon is observed herein, where the streamwise energy is recovered reasonably well whereas the predictive performance of wall-normal and spanwise energy and the pressure perturbation is degraded especially near the wall, when only the error in streamwise energy is minimized.

On the other hand, when $(\gamma_{uu}, \gamma_{vv}, \gamma_{ww}, \gamma_{pp})=(1,1,1,1)$ is used in the loss function, the prediction results are shown in figure \ref{testing2}. Only the case in the testing process when $Re_\tau=1000$ is presented herein, for brevity. It is found that, in contrast to the results in figure \ref{testing1}, the predictive performance of $E_{vv}^+$, $E_{ww}^+$ and $E_{pp}^+$ is improved, to some extent, at the expense of the reducing the predictive accuracy of $E_{uu}^+$, similar to the performance in ref. \cite{Moarref2013a}. A slight mismatch of the inner peak of $E_{uu}^+$ appears in panel ($a$).
As we mentioned earlier, this might be attributed to the ratio of the streamwise response to the wall-normal, spanwise and pressure response inherent in the rank-1 model. In this case, the matching error of the velocities and pressure cannot be simultaneously small enough, which means that the rank-1 approximation by optimal response modes might lead to insufficient basis functions for modelling.
Nonetheless, the predictive performance presented here is still believed to be reasonably acceptable.

To be more quantitative, Table \ref{tab1} lists the results of MSE of prediction, in terms of each component. Consistent with the qualitative description above, both the loss function used can yield acceptable performance in prediction, e.g. $MSE\le10\%$, except for the cases at very-low Reynolds numbers where the inner-outer characteristics are not 
remarkable and can be a bit different from those in the higher-Reynolds-number cases.
Specifically, the choice of the loss function can be determined by the prediction target.

\begin{center}
\begin{table}[h]
\caption{MSE of each component of wall-normal energy in the training and testing process.}\label{tab1}
\resizebox{\linewidth}{!}{
\centering
\begin{tabular}{@{}*{10}{c}}
\br
\multicolumn{2}{c}{\multirow{2}{*}{$Re_\tau$}} & \multicolumn{4}{c}{$(\gamma_{uu}, \gamma_{vv}, \gamma_{ww}, \gamma_{pp})=(1,0,0,0)$} & \multicolumn{4}{c}{$(\gamma_{uu}, \gamma_{vv}, \gamma_{ww}, \gamma_{pp})=(1,1,1,1)$}  \\
 & & $MSE_{uu}$(\%)& $MSE_{vv}$(\%)& $MSE_{ww}$(\%)& $MSE_{pp}$(\%)& $MSE_{uu}$(\%)& $MSE_{vv}$(\%)& $MSE_{ww}$(\%)& $MSE_{pp}$(\%) \\  \mr
\multirow{6}{*}{Training}&180 & 2.32 & 66.78 & 39.45 & 56.67 & 5.29 & 53.78 & 29.95 & 10.17 \\
&380 & 0.62 & 11.61 & 7.56 & 7.97 & 1.92 & 8.54 & 4.66 & 0.85 \\
&550 & 0.29 & 6.37 & 4.28 & 7.33 & 1.13 & 5.32 & 2.90 & 0.90  \\
&2000 & 0.05 & 4.38 & 3.08 & 2.71 & 4.83 & 3.71 & 2.49 & 2.07 \\
&4200 & 0.19 & 6.97 & 6.66 & 4.90 & 0.10 & 5.89 & 2.76 & 2.62 \\
&5200 & 0.16 & 8.03 & 8.19 & 4.85 & 1.09 & 6.59 & 2.92 & 2.30\\ \mr
Testing &1000 & 0.44 & 3.27 & 2.08 & 2.64 & 4.72 & 2.69 & 1.40 & 1.46 \\\br
\end{tabular}}
\end{table}
\end{center}

\begin{figure}
\centering
\subfigure{\includegraphics[width=1.0\textwidth]{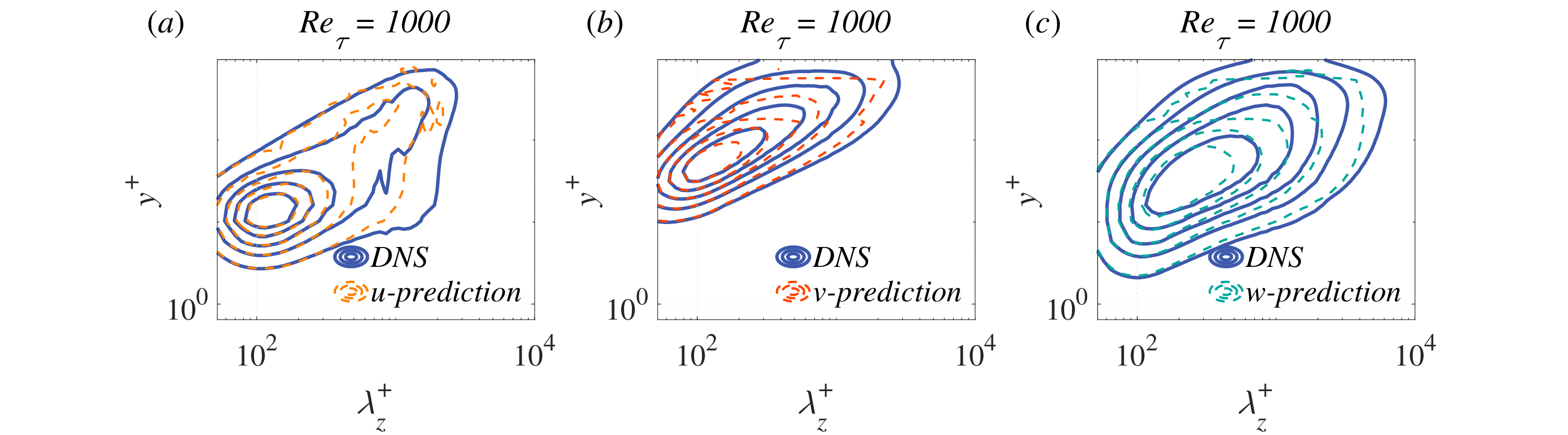}}
\caption{The prediction of spanwise spectra ($a$) $E_{uu}^+$, ($b$) $E_{vv}^+$ and ($c$) $E_{ww}^+$ in the testing process, when $(\gamma_{uu}, \gamma_{vv}, \gamma_{ww}, \gamma_{pp})=(1,0,0,0)$, in the channel flow at $Re_\tau=1000$. The contour lines represent 1/6, 1/3, 1/2, 2/3 and 5/6 of their respective maximum, moving inward.}\label{testing3}
\end{figure}

\begin{figure}
\centering
\subfigure{\includegraphics[width=1.0\textwidth]{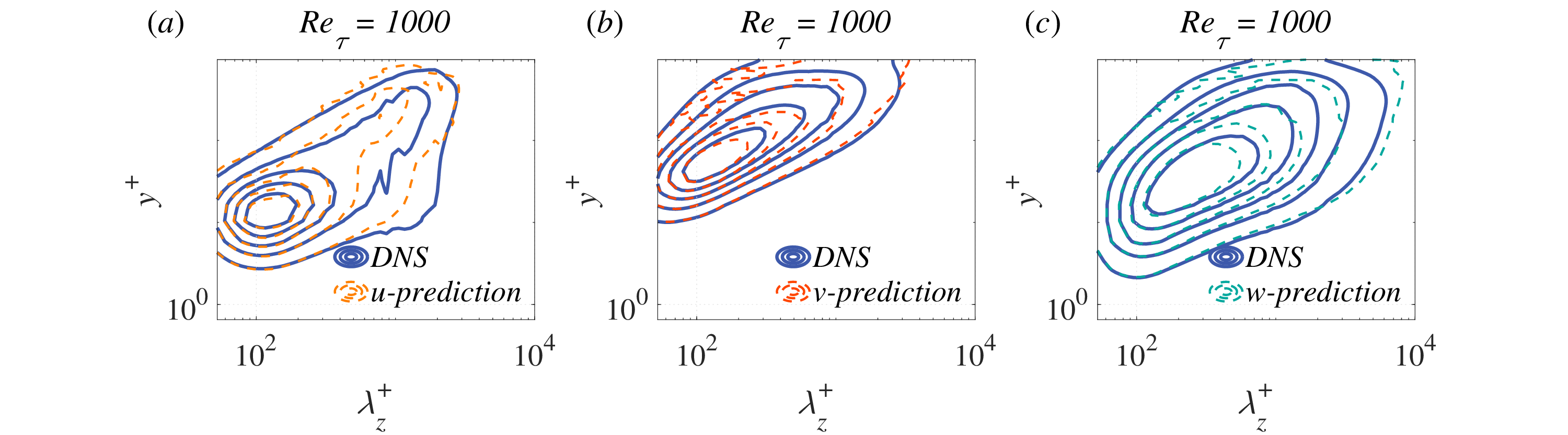}}
\caption{The prediction of spanwise spectra ($a$) $E_{uu}^+$, ($b$) $E_{vv}^+$ and ($c$) $E_{ww}^+$ in the testing process, when $(\gamma_{uu}, \gamma_{vv}, \gamma_{ww}, \gamma_{pp})=(1,1,1,1)$, in the channel flow at $Re_\tau=1000$.}\label{testing4}
\end{figure}

We also use the resolvent-based modelling \eqref{Eqq1} in combination with machine learning to predict the spanwise spectra of velocity.
The prediction results in the testing process are shown in figures \ref{testing3} and \ref{testing4}, for the cases when  $(\gamma_{uu}, \gamma_{vv}, \gamma_{ww}, \gamma_{pp})=(1,0,0,0)$ and $(\gamma_{uu}, \gamma_{vv}, \gamma_{ww}, \gamma_{pp})=(1,1,1,1)$, respectively. 
In both cases, the inner-layer peaks in the premultiplied spanwise spectra, $E_{uu}^+$, $E_{vv}^+$ and $E_{ww}^+$, which are representative of the near-wall streaks and streamwise vortices, can be accurately captured.  Whereas for larger scales in the outer-layer cycle, the prediction error increases, which is possibly due to the limitation of the databases used for training as the outer-layer characteristics exist evidently only at high Reynolds numbers, that is $Re_\tau = 2000$ and $Re_\tau = 5200$ in the present study.
Similar to the discussion above, the loss function with $(\gamma_{uu}, \gamma_{vv}, \gamma_{ww}, \gamma_{pp})=(1,1,1,1)$ is able to increase the prediction ability of $E_{vv}^+$ and $E_{ww}^+$, whereas decrease that of $E_{uu}^+$, which is straightforwardly exposed by the MSE of each component listed in Table \ref{tab2}. 
In general, the streamwise energy spectra can be predicted well, as $MSE_{uu}<6\%$ for all the cases, while the $MSEs$ of the wall-normal and spanwise energy spectra  are much larger.
Hence the full-field $a_1$ or higher-order models might be necessary and more databases for training and testing will definitely improve the performance of turbulence prediction.

\begin{center}
\begin{table}[h]
\caption{MSE of each component of energy spectra in the training and testing process.}\label{tab2}
\resizebox{\linewidth}{!}{
\centering
\begin{tabular}{@{}*{8}{c}}
\br
\multicolumn{2}{c}{\multirow{2}{*}{$Re_\tau$}} & \multicolumn{3}{c}{$(\gamma_{uu}, \gamma_{vv}, \gamma_{ww}, \gamma_{pp})=(1,0,0,0)$} & \multicolumn{3}{c}{$(\gamma_{uu}, \gamma_{vv}, \gamma_{ww}, \gamma_{pp})=(1,1,1,1)$}  \\
 & & $MSE_{uu}$(\%)& $MSE_{vv}$(\%)& $MSE_{ww}$(\%)& $MSE_{uu}$(\%)& $MSE_{vv}$(\%)& $MSE_{ww}$(\%) \\  \mr
\multirow{4}{*}{Training} &180 & 1.4945 & 10.5316 & 11.3235 & 4.2361 & 13.9983 & 20.0548\\
 &550 & 1.4359 & 2.3791 & 9.9905 & 3.0671 & 2.1305 & 4.0066  \\
 &2000 & 0.72158 & 25.1527 & 20.2006 & 1.1954 & 13.4671 &9.146\\
 &5200&4.1169& 51.7848 & 45.8747 & 1.5947 & 33.8016 & 26.9096\\
 \mr
Testing &1000 & 5.941 & 10.5823 & 7.2193 & 5.0774 & 4.7236 &3.2054\\
\br
\end{tabular}}
\end{table}
\end{center}

\section{Conclusion}\label{conclusion}
This paper proposes a modelling framework based on the resolvent analysis and machine learning, and uses it to predict the turbulent energy in incompressible channel flows.
In this method, the mean streamwise velocity profiles are required as input to the network. On one hand, the input works to provide efficient optimal response modes as basis functions, since the low-rank behaviour of the resolvent operator is verified. On the other hand, the projection coefficients of the low-rank forcing modes onto the true nonlinear terms are obtained through a CNN architecture, adjusting the magnitude of the linearly superposed response at each wavenumber combination and consequently leading to the reconstruction of the flow perturbations.

To optimize the weight and bias parameters in neurons per hidden layer during the training process, two types of loss functions are considered, with matching error of different components included.
Results show that both the methods can yield a reasonably good prediction of the wall-normal profile of turbulent fluctuations, except that better performance is observed in the prediction of the streamwise component when only the matching error of the streamwise energy is penalized.
As for the prediction of spanwise spectra of velocity, the inner-layer peaks can be accurately captured whereas the prediction error increases for larger scales in the outer-layer cycle. 
In general, the proposed method provides a promising tool to predict the turbulent motions based on the mean field obtained from low-cost RANS or experimental measurements. Further work will be undertaken to investigate the effect of increasing the degrees of freedom of the projection coefficient and the numbers of resolvent-based modes on the predictive accuracy.

\section*{Acknowledgments}

The funding support of the National Natural Science Foundation of China (under the Grant
Nos. 91952302 and 92052101) is acknowledged.
This work was supported in part by the European Research Council under the Caust grant ERC-AdG-101018287. The authors also thank Prof. Ricardo Vinuesa for his suggestions and comments on this manuscript.

\section*{References}
\bibliographystyle{iopart-num}
\bibliography{./reference}

\end{document}